\documentclass[a4paper,aps,prb,twocolumn,english,superscriptaddress,floatfix,longbibliography]{revtex4-2}

\usepackage{graphicx}
\usepackage{dcolumn}
\usepackage{bm}
\usepackage{amsmath} 
\usepackage{color}
\usepackage{pifont}
\usepackage{natbib}
\usepackage{siunitx}
\usepackage{soul}
\usepackage{verbatim}
\usepackage{placeins}
\usepackage{hyperref}
\usepackage{cleveref}

\hypersetup{
colorlinks = true,
urlcolor   = blue,
linkcolor  = blue,
citecolor  = blue
}

\usepackage{float}
\DeclareUnicodeCharacter{2212}{\textminus}

\begin{document}

\title{Microscopic magnetic phase evolution in the Weyl semimetal $\mathrm{Mn}_3\mathrm{Sn}$ revealed by $\mu^+\mathrm{SR}$}

\author{U.~Miniotaite}
 \email{ugnem@kth.se}
\affiliation{Department of Applied Physics, KTH Royal Institute of Technology, SE-106 91 Stockholm, Sweden}

\author{O.~K.~Forslund}
\affiliation{Department of Physics and Astronomy, Uppsala University, Uppsala, Sweden}

\author{H.~Luetkens}
\affiliation{PSI Center for Neutron and Muon Sciences CNM, Villigen PSI, Switzerland}

\author{V.~Rai}
\affiliation{Department of Physics and Materials Science, University of Luxembourg, 162A Avenue de la Faiencerie, 1511 Luxembourg, Grand Duchy of Luxembourg}
\affiliation{Forschungszentrum J\"ulich GmbH, J\"ulich Centre for Neutron Science (JCNS-2) and Peter Gr\"unberg Institut (PGI-4), JARA-FIT, 52425 J\"ulich, Germany}

\author{J.~Per{\ss}on}
\affiliation{Forschungszentrum J\"ulich GmbH, J\"ulich Centre for Neutron Science (JCNS-2) and Peter Gr\"unberg Institut (PGI-4), JARA-FIT, 52425 J\"ulich, Germany}

\author{G.~D.~Morris}
\affiliation{Centre for Molecular and Materials Science, TRIUMF, Vancouver, British Columbia, Canada V6T 2A3}

\author{D.~Zuhair}
\affiliation{Department of Applied Physics, KTH Royal Institute of Technology, SE-106 91 Stockholm, Sweden}

\author{D.~W.~Tam}
\affiliation{Department of Applied Physics, KTH Royal Institute of Technology, SE-106 91 Stockholm, Sweden}
\affiliation{Institut Laue-Langevin, CS 20156, 38042 Grenoble C\'edex 9, France}

\author{S.~Nandi}
\affiliation{Forschungszentrum J\"ulich GmbH, J\"ulich Centre for Neutron Science (JCNS-2) and Peter Gr\"unberg Institut (PGI-4), JARA-FIT, 52425 J\"ulich, Germany}
\affiliation{RWTH Aachen, Lehrstuhl f\"ur Experimentalphysik IVc, J\"ulich-Aachen Research Alliance (JARA-FIT), 52074 Aachen, Germany}

\author{Y.~Sassa}
 \email{sassa@kth.se}
\affiliation{Department of Applied Physics, KTH Royal Institute of Technology, SE-106 91 Stockholm, Sweden}

\author{M.~M{\aa}nsson}
 \email{condmat@kth.se}
\affiliation{Department of Applied Physics, KTH Royal Institute of Technology, SE-106 91 Stockholm, Sweden}

\begin{abstract}
\noindent We report a comprehensive muon spin relaxation ($\mu^+\mathrm{SR}$) and bulk magnetization study of the antiferromagnetic (AFM) Weyl semimetal $\mathrm{Mn}_3\mathrm{Sn}$ (composition $\mathrm{Mn}_{2.99}\mathrm{Sn}$). $\mathrm{Mn}_3\mathrm{Sn}$ is reported to exhibit a commensurate inverse triangular (IT) AFM phase, an incommensurate (IC) helical AFM phase, and a proposed low-temperature spin-glass-like state. In our sample, we identify the characteristic temperatures associated with these regimes as N\'eel temperature ($T_\mathrm{N} = 418$~K), a macroscopic bulk transition temperature between IT-AFM to IC helical phase ($T_\mathrm{t} \approx 275$~K), and low-temperature transition $T_\mathrm{f} = 21$~K. Investigating the low-temperature regime below $T_\mathrm{f}$, we find no evidence of a static spin-glass state. Instead, the sample exhibits an increasing ferromagnetic (FM) component accompanied by a localized slowing of spin fluctuations, indicating that these phenomena may be decoupled. In the IC helical AFM phase, the zero-field (ZF) spectra are best described by damped oscillations with an empirical phase offset, consistent with anharmonic and amplitude-modulated order reported by scattering studies. Upon warming above 150~K, a continuous redistribution of muon spectral weight reveals a broad, homogeneous magnetic crossover between the IC helical and IT-AFM phases. In the commensurate IT-AFM phase above $T_\mathrm{t}$, a persistent missing fraction in the initial asymmetry indicates that a subset of implanted muons, corresponding to roughly 20\% of the sample-related asymmetry, undergoes unresolved ultrafast depolarization. Finally, we observe temperature-driven shifts in muon site populations above 325~K. Ultimately, our results show a highly dynamic magnetic landscape in $\mathrm{Mn}_3\mathrm{Sn}$, demonstrating how its complex magnetic orders often coexist and evolve continuously with temperature.
\end{abstract}

\maketitle

\section{Introduction}
\noindent Topological materials have emerged as a central theme in condensed matter physics, in large part because their non-trivial band topology can generate unconventional transport responses that are useful for spintronics and low-power electronics~\cite{Nadeem2021,Tian2017,Smejkal2018_TopoAFMSpintronics}. Among these, antiferromagnetic (AFM) Weyl semimetals (WSMs) have been especially appealing. They combine the robustness and ultrafast dynamics of AFMs with anomalously large, Berry-curvature-driven signals exceeding those of conventional AFMs~\cite{Hasan2010_TI,Yan2017_TopoMaterials,Smejkal2018_TopoAFMSpintronics}. A prototypical example is Mn$_3$Sn, for which experimental evidence of magnetic Weyl fermions has been reported and which displays a large anomalous Hall effect at room temperature despite having a very small net magnetization~\cite{Nakatsuji2015_AHE,Kuroda2017_MagneticWeyl,Chen2021_Mn3X}.

Mn$_3$Sn crystallizes in a hexagonal structure (space group P$6_3/mmc$) where Mn atoms form a kagome lattice in the $ab$-plane. At room temperature, Mn$_3$Sn adopts an inverse triangular AFM (IT-AFM) structure with a small uncompensated ferromagnetic (FM) moment~\cite{Mn3Sn_struc1,Sandratskii1996_WeakFM,Kren1975_Mn3SnPhaseTrans,Mn3Sn_struc2,Tomiyoshi1982_MagStructWeakFM,Nagamiya1982}. When cooled below $T_\mathrm{t}\approx 275~\mathrm{K}$, with reported values strongly dependent on synthesis conditions and sample stoichiometry ($\sim 200$--290~K)~\cite{Deng2022_ResidualStrainMn3Sn,Song2020_ComplicatedMagStruct,Jacobsen_2026,Tomiyoshi1987,Park2025_NominalMn3Sn}, it enters a complex helical AFM state. Early neutron scattering studies already identified multiple incommensurate (IC) propagation vectors along the $c$-axis (e.g., $k_z^1\approx 0.08$ and $k_z^2 \approx 0.10$)~\cite{Cable1993,Park2018,Mn3Sn_struc2}. More recent neutron and X-ray studies further show that the low-temperature state is not a uniform helical spiral, but involves amplitude-modulated spin-density-wave/helical components and intertwined spin and charge density waves~\cite{Wang2023multiK,Chen2024_IntertwinedCDWSDW}.

\begin{figure*}[!ht]
\centering 
\includegraphics[width = 0.8\textwidth]{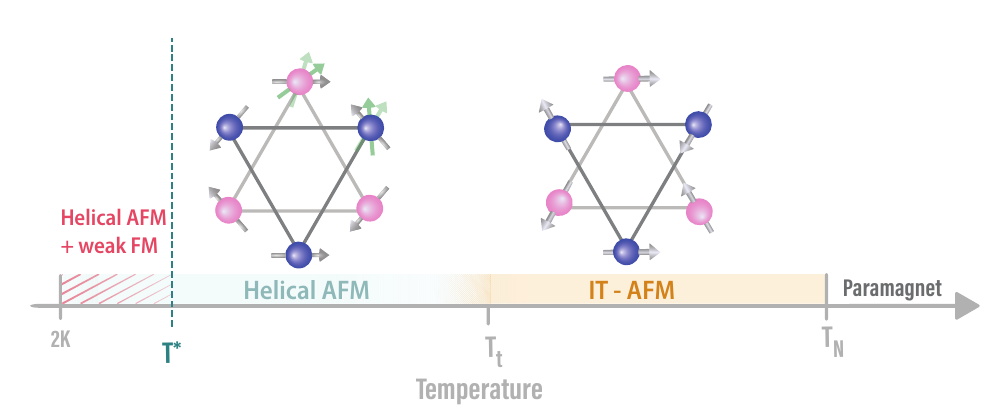}
\caption{\textbf{Magnetic phase diagram of Mn$_{3}$Sn.} The evolution of magnetic states is shown as a function of temperature, indicating key transition temperatures ($T_\mathrm{f}$, $T_\mathrm{t}$, and $T_\mathrm{N}$). Exact values are omitted here because transition temperatures vary between samples. The spin configurations of the Mn atoms on the kagome lattice are illustrated for the helical antiferromagnetic (AFM) and inverse triangular AFM (IT-AFM) phases. Blue and pink atoms denote Mn atoms located in different layers in the unit cell. Green arrows show the IC spin rotation in subsequent unit cells for the respective layers.}
\label{fig:magstruc}
\end{figure*}

Upon further cooling below $T_{\mathrm{f}}$ (typically 30~K--50~K),  Mn$_3$Sn exhibits a magnetic transition characterized by a strong bifurcation between zero-field-cooled (ZFC) and field-cooled (FC) magnetization measurements \cite{Feng2006,Song2020_ComplicatedMagStruct, Ohmori1987}. Based on the bifurcation and a small frequency-dependent shift in $\chi'$ in AC magnetic susceptibility (ACMS) measurements, studies have attributed this regime to a partial FM spin-glass state \cite{Feng2006, Deng2022_ResidualStrainMn3Sn}, with speculation regarding its origin coming from excess Mn atoms \cite{Feng2006, Ikhlas2020}. These three characteristic magnetic transitions are illustrated in Fig. \ref{fig:magstruc}.

Despite extensive bulk and neutron scattering studies, the microscopic nature of the low-temperature anomalous phase and the evolution from the IC helical state into the IT-AFM state remain unresolved. For example, neutron diffraction has not detected a clear change in the long-range AFM order across $T_\mathrm{f}$~\cite{Song2020_ComplicatedMagStruct, Tomiyoshi1986_TriangularSpin}. Furthermore, since scattering and bulk magnetization experiments measure a macroscopic average, details of the local magnetic environment, such as spin dynamics and subtle phase coexistence, cannot be definitively determined. This makes the system an ideal target for muon spin relaxation ($\mu^+\mathrm{SR}$), a local probe highly sensitive to internal magnetic fields and spin dynamics.

Here, we resolve these uncertainties for the nearly stoichiometric $\mathrm{Mn}_{2.99}\mathrm{Sn}$ by combining $\mu^+\mathrm{SR}$ with bulk magnetization measurements.  Tracking the temperature evolution of the muon depolarization reveals a broad redistribution of fitted oscillatory amplitudes that begins near 150~K and extends to the bulk transition temperature $T_\mathrm{t} \approx 275$~K, indicating a local magnetic crossover from the IC helical phase to the IT-AFM regime. Furthermore, when examining the low-temperature transition across $T_{\mathrm{f}}$, we find no clear $\mu^+\mathrm{SR}$ signature of static freezing. Instead, we observe an increasing FM component accompanied only by a localized slowing of spin fluctuations. This indicates that the low-temperature FM response in this sample is decoupled from the glassy freezing found in some $\mathrm{Mn}_3\mathrm{Sn}$ samples~\cite{Feng2006}. Ultimately, our results map a highly dynamic magnetic landscape, demonstrating that the complex magnetic orders in $\mathrm{Mn}_3\mathrm{Sn}$ do not undergo abrupt phase transitions, but instead coexist and evolve through a continuous cross-over with temperature.

\section{Experimental Methods}
While $\mathrm{Mn}_3\mathrm{Sn}$ is often stabilized with an excess of Mn (resulting in $\mathrm{Mn}_{3+x}\mathrm{Sn}_{1-x}$), and its magnetic response is sensitive to excess Mn and sample quality~\cite{Ikhlas2020,Park2025_NominalMn3Sn}, the sample used in this study was specifically synthesized to closely approach ideal stoichiometry~\cite{Kren1975_Mn3SnPhaseTrans}. Mn and Sn were melted in a stoichiometric ratio of 3:1 using a self-flux growth method. This process was carried out in an argon atmosphere to ensure homogeneous mixing of the two elements. The resulting alloy was then sealed in a quartz tube under an argon atmosphere and heated to 1273~K for 12~h. Subsequently, the sample was cooled at a rate of 2~K/h until it reached 973~K. Finally, the sample was quenched in water to rapidly cool it to room temperature. The exact chemical composition of the sample was found to be $\mathrm{Mn}_{2.99}\mathrm{Sn}$ using inductively coupled plasma optical emission spectroscopy (ICP-OES). The crystal was ground to a fine powder using an agate mortar and pestle. For simplicity, we refer to this sample as $\mathrm{Mn}_3\mathrm{Sn}$ throughout the remainder of this paper.

Magnetic measurements were performed at the AlbaNova NanoLab using a Quantum Design PPMS DynaCool. A 32~mg powder sample was compacted in a capsule. Magnetization (VSM option) was measured from 5~K to 380~K at 1000~Oe, with isothermal $M(H)$ curves collected between 5~K and 300~K in fields up to 5~T. AC susceptibility (ACMS II option) was acquired between 500~Hz and 10~kHz from 5~K to 80~K using a 5~Oe oscillating field.

The $\mu^+\mathrm{SR}$ experiment was conducted on the General Purpose Surface-Muon Beamline (GPS) at the Paul Scherrer Institute (PSI) in Villigen, Switzerland. Approximately \SI{400}{\milli\gram} of fine powder was pressed into a pellet (10~mm diameter) to avoid reorientation of the crystallites due to magnetic ordering or an applied external magnetic field. The pellet was pressed with a small amount of Apiezon-H grease to ensure thermal conductivity above room temperature. The sample was then glued to a low-background copper film using GE varnish for thermal contact with the heater, mounted on a low-background copper fork, and inserted into a CCR cryostat, allowing measurements from \SI{5}{\kelvin} to \SI{475}{\kelvin}.

The $\mu^+\mathrm{SR}$ measurements were performed in different magnetic field configurations, defined by the field alignment relative to the initial muon spin polarization: zero-field (ZF) and transverse field (TF). In ZF, no external field is applied, and in TF, the field is perpendicular to the initial muon spin polarization.

ZF and TF spectra were collected for $\mathrm{Mn}_3\mathrm{Sn}$ between \SI{5}{K} and \SI{475}{K}. For a set of selected temperatures (5~K, 250~K, 300~K, and 350~K), higher-statistics ZF measurements were conducted. TF measurements were performed using a magnetic field of $B = \SI{50}{G}$. The data were analyzed using the \texttt{musrfit} software~\cite{musrfit}.

\section{MAGNETIZATION RESULTS}\label{sec:mag_all}

\begin{figure}[htbp]
    \centering
    \includegraphics[width=1.\linewidth]{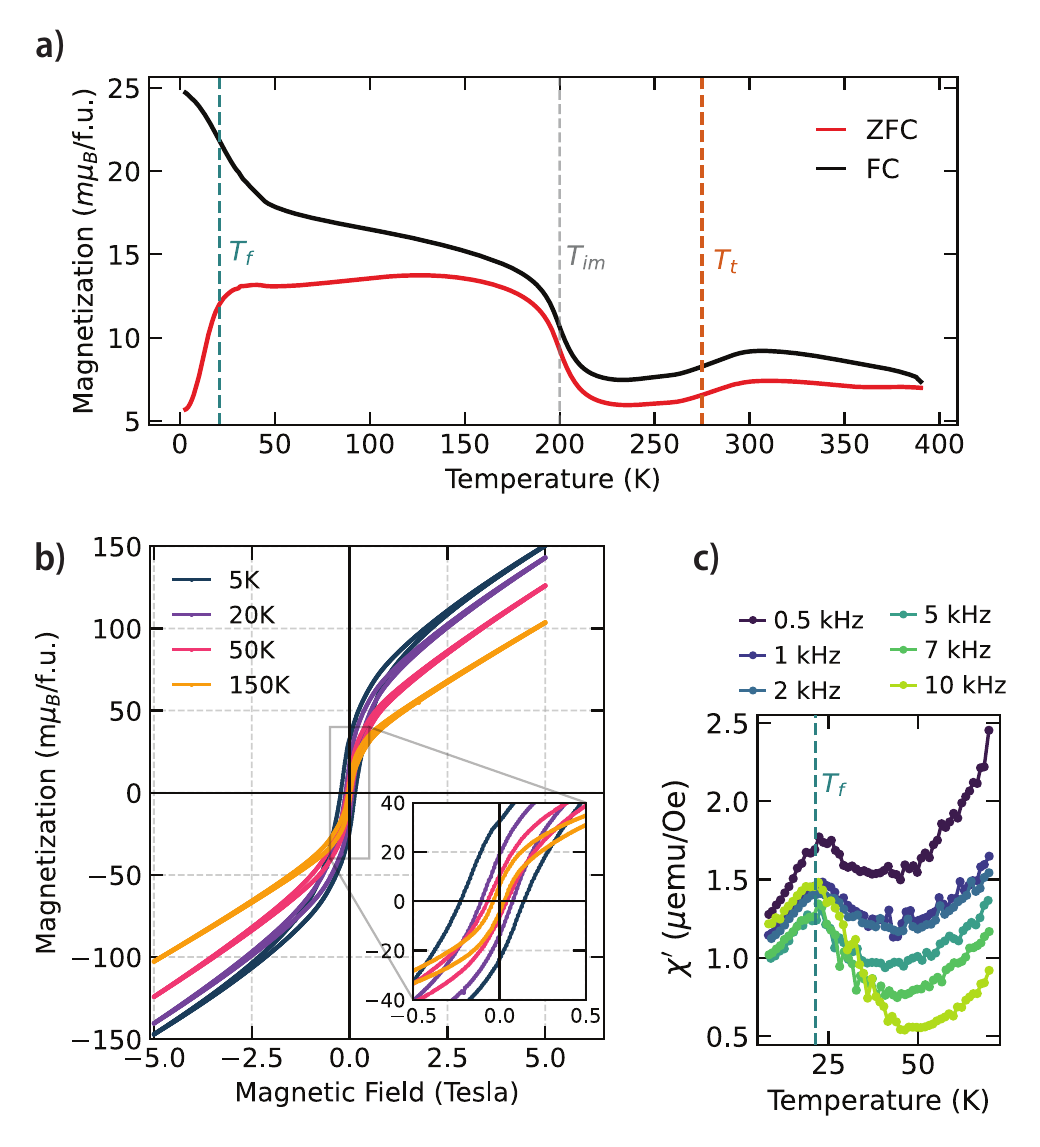}
    \caption{\textbf{Bulk magnetic characterization.} \textbf{a)} Temperature dependence of the magnetization using zero-field-cooled (ZFC) and field-cooled (FC) protocols with an applied magnetic field of $B = 0.1$~T, marking key transition temperatures ($T_\mathrm{f}$, $T_\mathrm{im}$, and $T_\mathrm{t}$). \textbf{b)} $M(H)$ curves showing magnetization as a function of applied magnetic field between -5~T and 5~T at 5~K, 20~K, 50~K, and 150~K. The inset provides a magnified view of the low-field region to highlight the temperature evolution of the hysteresis. \textbf{c)} Real part of the AC magnetic susceptibility (ACMS), $\chi'$, as a function of temperature across various frequencies, showing a peak near $T_\mathrm{f}$.}
    \label{fig:Mag_all}
\end{figure}

Fig.~\ref{fig:Mag_all}(a) shows the temperature dependence of magnetization, $M(T)$, for both zero-field-cooled (ZFC) and field-cooled (FC) protocols for $\mathrm{Mn}_3\mathrm{Sn}$ measured during heating under an applied magnetic field of $B = 0.1$~T. From the derivative of the magnetization curve, we identify three clear macroscopic transitions: $T_\mathrm{f}~=~21$~K, $T_\mathrm{im} = 200$~K, and $T_\mathrm{t}~=~275$~K. The transition at $T_\mathrm{im} = 200$~K is attributed to a magnetic impurity. The transition across $T_\mathrm{t}~=~275$~K is approximately 20~K broad. Additionally, below $T_\mathrm{f}$, we observe a strong bifurcation where the ZFC and FC curves diverge.

Fig.~\ref{fig:Mag_all}(b) shows the $M(H)$ curves measured at 5~K, 20~K, 50~K, and 150~K. At 150~K, the $M(H)$ loop exhibits a sharp S-shape with small hysteresis. Between 150~K and 50~K, the width of the hysteresis loop is mostly unchanged, but the total magnetization slope increases. This increase coincides with the widening bifurcation observed in the $M(T)$ data [Fig.~\ref{fig:Mag_all}(a)]. As the temperature is further reduced to 20~K and 5~K, the hysteresis increases from approximately 0.1~T at 50~K to 0.35~T at 5~K. A complete set of $M(H)$ curves over a broader temperature range is available in Appendix~\ref{SI:mag}.

Fig.~\ref{fig:Mag_all}(c) shows the real part of the ACMS of $\mathrm{Mn}_3\mathrm{Sn}$ between 5~K and 80~K. A distinct peak is observed at $T_\mathrm{f} = 21$~K. Notably, our results do not show a frequency-dependent shift for this peak.

\section{MUON SPIN SPECTROSCOPY RESULTS}\label{sec:ZF}

\subsection{Transverse Field }\label{sec:TF}
\begin{figure*}[ht!]
    \centering
    \includegraphics[width = 1\textwidth]{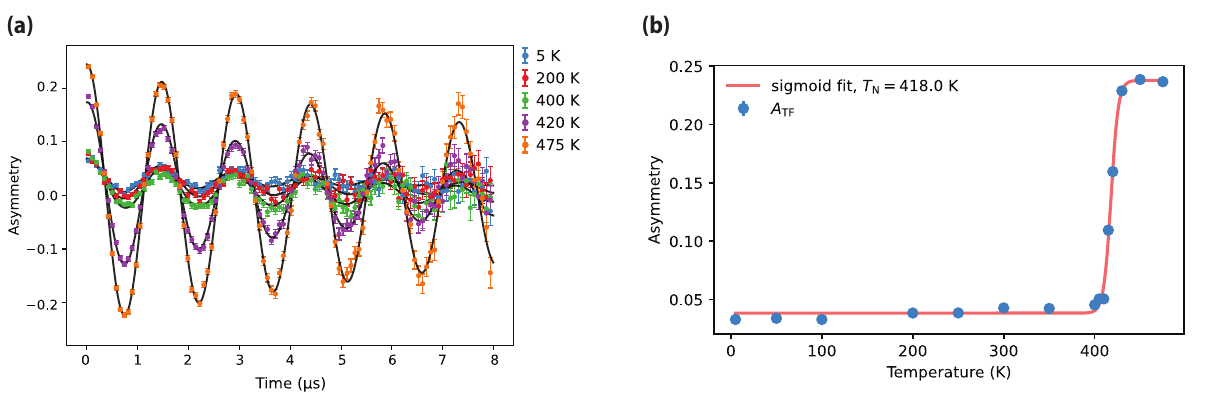}
    \caption{\textbf{N\'eel temperature from TF-$\boldsymbol{\mu}^+$\textbf{SR}.} \textbf{(a)} Muon spin relaxation ($\mu^+\mathrm{SR}$) spectra of $\mathrm{Mn}_3\mathrm{Sn}$ measured in an applied transverse field (TF) of $B = \SI{50}{G}$ at various temperatures. Solid black lines show fits to the experimental data using Eq.~\eqref{eq:wTF_polar}. \textbf{(b)} Evolution of the fitted TF asymmetry coefficient, $A_{TF}$, as a function of temperature. The sharp magnetic transition is fitted with a sigmoid function (red curve), extracting a N\'eel temperature of $T_\mathrm{N} = \SI{418}{K}$.}
    \label{fig:wTF_data}
\end{figure*}

To measure the N\'eel temperature and estimate possible minority/background contributions, we first conducted $\mu^+\mathrm{SR}$ measurements in an applied transverse field (TF) of $B = \SI{50}{G}$. The complete muon depolarization spectra for selected temperatures are shown in Fig.~\ref{fig:wTF_data}(a). The muon polarization function used to fit the TF data is described by:

\begin{equation} \label{eq:wTF_polar}
\begin{split}
    {A_\mathrm{tot} P_{\mathrm{TF}}(t)} =& {A_{\mathrm{TF}} \cos{\left(2 \pi f_\mathrm{TF} t + \frac{\pi \phi_\mathrm{TF}}{180}\right)} e^{- \lambda_{\mathrm{TF}} t}}\\
     +& A_{\mathrm{tail}}\,e^{- \lambda_{\mathrm{tail}} t}, 
\end{split}
\end{equation}
where $P_\mathrm{TF}(t)$ is the muon polarization in the TF, $A_\mathrm{tot}$ is the total fitted asymmetry, $A_\mathrm{TF}$ describes the relaxed oscillating component from the applied external field, and $A_\mathrm{tail}$ is the slowly decaying magnetic tail arising from internal magnetic moments parallel to the muon spin. The parameter $f_\mathrm{TF}$ is the oscillation frequency for the TF, while $\lambda_\mathrm{TF}$ and $\lambda_\mathrm{tail}$ are the respective relaxation rates.

At low temperatures, the TF asymmetry is strongly suppressed by the static internal magnetic ordering of the sample. Upon heating past $T_\mathrm{N}$ into the paramagnetic phase, the full asymmetry of the TF oscillations is recovered. By fitting the temperature dependence of the TF asymmetry with a sigmoid function, we extract a N\'eel temperature of $T_\mathrm{N} = 418$~K for $\mathrm{Mn}_3\mathrm{Sn}$ [Fig.~\ref{fig:wTF_data}(b)].The instrumental asymmetry calibration parameter ($\alpha$) was determined from the fully paramagnetic state at 450~K and kept fixed throughout the analysis.

Below $T_\mathrm{N}$, the TF asymmetry is not completely suppressed, remaining at a baseline of $\approx 13\%$ ($A_\mathrm{TF} \approx 0.033$) at 5~K. A large part of this residual signal is expected to arise from muons stopping outside the sample, including in the Cu sample holder and protective Cu foil, as well as in the grease and GE varnish used to mount the powder pellet. This baseline value is therefore used to constrain the paramagnetic background contribution in the ZF fits.

In addition to this background, $A_\mathrm{TF}$ exhibits two small recoveries below $T_\mathrm{N}$, each corresponding to only $\sim 2\%$ of the total instrumental asymmetry. The first recovery, between approximately 100~K and 200~K, occurs near the impurity-related feature observed in the bulk magnetization and is consistent with a small magnetic impurity fraction. The second recovery, appearing above approximately 250~K, occurs close to the IC-helical to IT-AFM transition and should not be assigned unambiguously to an impurity phase from the TF data alone. It may instead reflect a change in the internal-field distribution of the main $\mathrm{Mn}_3\mathrm{Sn}$ phase, for example through partial field cancellation at a minority muon stopping environment as the magnetic structure becomes more symmetric. We cannot discount that this shift could originate from small changes in $\alpha$ caused by thermal expansion of the sample and holder over this large temperature range. However, the magnitude of these recoveries remains very small, setting an upper bound on possible minority contributions and confirming that the dominant ZF-$\mu^+\mathrm{SR}$ signal originates from the primary $\mathrm{Mn}_3\mathrm{Sn}$ phase.

\subsection{Zero Field}
\begin{figure}[!ht]
    \centering
    \includegraphics[width = 0.45\textwidth]{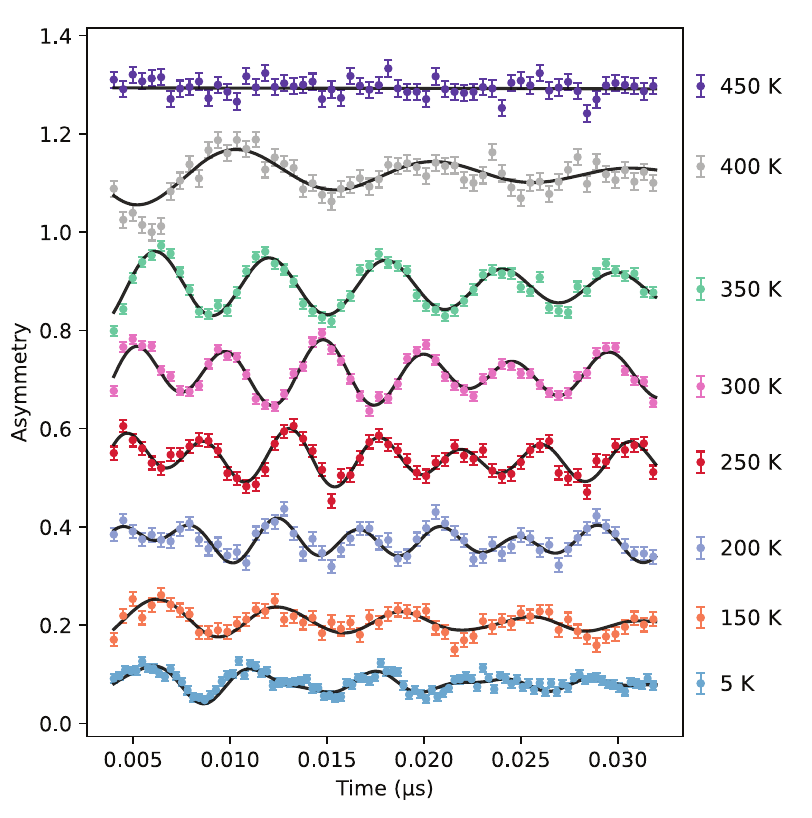}
    \caption{\textbf{Temperature evolution of ZF-$\boldsymbol{\mu}^+$\textbf{SR} spectra.} Zero-field (ZF) muon spin relaxation ($\mu^+\mathrm{SR}$) asymmetry of $\mathrm{Mn}_3\mathrm{Sn}$ measured at various temperatures from 5~K to 450~K. Solid black lines show fits to the experimental data using Eq.~\eqref{eq:ZF}. Successive spectra are vertically offset by a constant value for clarity of presentation.}
    \label{fig:ZF_spectra}
\end{figure}

The ZF $\mu^+\mathrm{SR}$ time spectra of $\mathrm{Mn}_3\mathrm{Sn}$ [Fig.~\ref{fig:ZF_spectra}] reveal a complex temperature-dependent evolution of rapid oscillations. The presence of oscillations indicates magnetic ordering, and multiple oscillations suggest magnetically inequivalent local-field components. For an IC magnetic structure, these do not need not correspond to crystallographically distinct muon stopping sites. In a powder, two-thirds of the internal magnetic-field components are expected to be perpendicular to the initial muon spin polarization (causing oscillations), whereas one-third is expected to be parallel to it (causing an exponentially decaying tail).
 
As $\mathrm{Mn}_3\mathrm{Sn}$ evolves through multiple magnetic phases, we observe changes in the frequencies. As the temperature increases above $T_\mathrm{N}$, the oscillations evolve to a static Gaussian Kubo-Toyabe (KT) function multiplied by an exponential depolarization~\cite{Hayano1979}, describing muon spin relaxation in a system of randomly oriented static spins within the muon lifetime. Spin fluctuations can additionally produce a fast-relaxing contribution to the polarization function. To account for all these processes within the considered temperature range, the ZF data were fitted with the following polarization function:

\begin{equation} \label{eq:ZF}
    \begin{split}
        A_\mathrm{tot} P_{\mathrm{ZF}}(t) =& \sum_{i=1}^{3} A_{\mathrm{AF}i} \cos{\left(2 \pi f_{\mathrm{AF}i} t + \frac{\pi \phi}{180}\right)} e^{- \lambda_{\mathrm{AF}i} t} \\
        &+ A_{\mathrm{F}}e^{- \lambda_{\mathrm{F}} t} + A_{\mathrm{KT}}G^{\mathrm{SGKT}}(t, \Delta_{\mathrm{KT}})e^{- \lambda_{\mathrm{KT}} t} \\
        &+ A_{\mathrm{tail}}e^{- \lambda_{\mathrm{tail}} t} + A_{\mathrm{BG}}e^{- \lambda_{\mathrm{BG}} t}
    \end{split}
\end{equation}

\begin{figure*}[!ht]
    \centering
    \includegraphics[width = 1\textwidth]{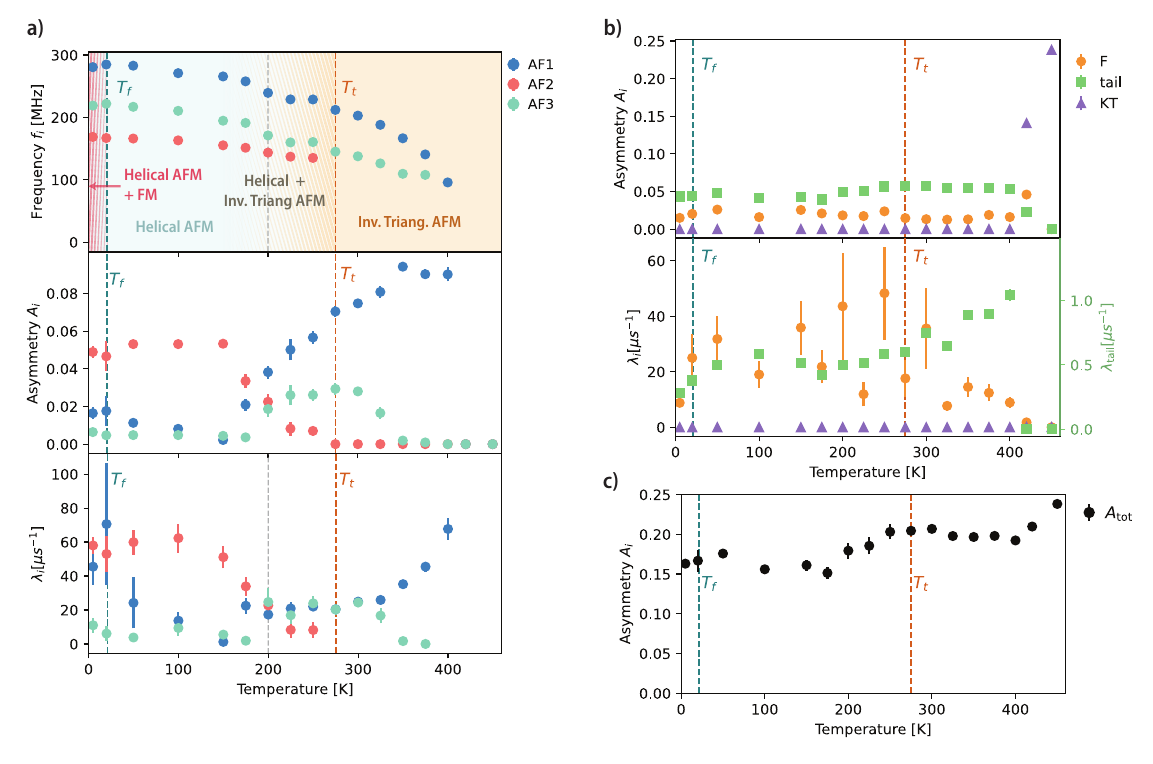}
    \caption{\textbf{Temperature dependence of the ZF-$\boldsymbol{\mu}^+$\textbf{SR} fit parameters.} Evolution of the parameters extracted from fitting the zero-field data to Eq.~\eqref{eq:ZF} for $\mathrm{Mn}_3\mathrm{Sn}$. Vertical dashed lines indicate the magnetic transition temperatures $T_f$ and $T_t$. \textbf{(a)} Fit parameters for the three oscillatory components (AF1, AF2, and AF3), showing the precession frequencies $f_i$ (top), asymmetries $A_i$ (middle), and relaxation rates $\lambda_i$ (bottom). Background shading corresponds to the magnetic phases defined in Fig.~\ref{fig:magstruc}. \textbf{(b)} Fit parameters for the non-oscillatory components, detailing the asymmetries (top) and relaxation rates (bottom) for the fast exponential (F), slow exponential tail (tail), and Kubo-Toyabe (KT) functions. \textbf{(c)} The total fitted initial asymmetry, $A_\mathrm{tot}$, as a function of temperature.}
    \label{fig:ZF_coefs}
\end{figure*}
where $P_\mathrm{ZF}(t)$ is the zero-field muon spin polarization and $A_\mathrm{tot}$ is the total fitted asymmetry. For the three oscillating AFM components ($i = 1, 2, 3$), $A_{\mathrm{AF}i}$ represents the asymmetry, $f_{\mathrm{AF}i}$ the precession frequency, $\lambda_{\mathrm{AF}i}$ the relaxation rate, and $\phi$ the shared initial phase. The Kubo-Toyabe (KT) contribution is described by a static Gaussian KT (SGKT) function, $G^{\mathrm{SGKT}}(t, \Delta_{\mathrm{KT}})$, with a distribution width $\Delta_\mathrm{KT}$, multiplied by an exponential relaxation rate $\lambda_\mathrm{KT}$. In addition to these, the polarization includes three purely exponential decay terms: a fast-relaxing contribution ($A_\mathrm{F}, \lambda_\mathrm{F}$) that may contain dynamic and/or unresolved static broadening, a very slowly decaying tail ($A_\mathrm{tail}, \lambda_\mathrm{tail}$) originating from internal magnetic fields parallel to the initial muon spin polarization, and a paramagnetic background signal ($A_\mathrm{BG}, \lambda_\mathrm{BG}$) established from transverse-field (TF) measurements (see Sec.~\ref{sec:TF}). In the fitting process, this background fraction, $A_\mathrm{BG}$, is fixed to the value determined from the base-temperature TF measurements. Furthermore, the phase of AF oscillations is theoretically $\phi = 0^\circ$. However, a finite fitted offset can arise from instrumental timing uncertainty as well as the broad, complex internal-field distributions characteristic of incommensurate magnetic order. We therefore treat $\phi$ as an empirical fitting parameter, constrained to be common to all frequencies.

\subsubsection{\texorpdfstring{{Low-temperature phase $\left( T \leq 50~\mathrm{K}\right)$}}{Low-temperature phase (T <= 50 K)}}

The temperature-dependent ZF fit parameters obtained from Eq.~\eqref{eq:ZF} are shown in Fig.~\ref{fig:ZF_coefs}. Around $T~=~20$~K, we observe a peak in the relaxation rate of the fastest oscillating component, $\lambda_{\mathrm{AF1}}$, coinciding with the transition, $T_\mathrm{f}$, observed in magnetization (see Sec.~\ref{sec:mag_all}). From a $\mu^+\mathrm{SR}$ perspective, an increase in the relaxation rate in an ordered magnet can reflect a broadening of static local-field distribution, a slowing of fluctuations into the $\mu^+\mathrm{SR}$ time window, or both~\cite{Dalmas97,Yaouanc2011}. Thus, although the macroscopic transition is not visible in the frequency or asymmetry parameters, we observe enhanced local-field relaxation in $\lambda_\mathrm{AF1}$ near $T_\mathrm{f}$. Upon cooling from $T = 50$~K to $T = 5$~K, the depolarization rate $\lambda_{\mathrm{F}}$ decreases from $\approx 35~\mu\mathrm{s}^{-1}$ to $10~\mu\mathrm{s}^{-1}$. At approximately constant $A_\mathrm{F}$, this decrease indicates that the corresponding fast-relaxing contribution becomes less effective at depolarizing the muon ensemble~\cite{Hayano1979,Yaouanc2011}. We also note that the relaxation parameters exhibit scatter and large error bars between 100~K and 300~K, and the apparently larger values often coincide with higher-statistics measurements. Ultimately, the small fitted amplitudes and fast relaxation of these components make them difficult to determine reliably across the entire temperature range.

\subsubsection{\texorpdfstring{{IC helical phase $\left( 5~\mathrm{K} < T < 275~\mathrm{K}\right)$}}{IC helical phase (5 K < T < 275 K)}}

At base temperature (5~K), three high-frequency oscillations are observed (the Fourier transform of the time spectra detailing these frequency components can be found in Appendix~\ref{app:FT}). We note that the total fitted asymmetry here, $A_\mathrm{tot} \approx 0.17$, is noticeably lower than the expected full instrumental asymmetry of $A_0 \approx 0.25$ [Fig.~\ref{fig:ZF_coefs}(c)]. In the low-temperature IC helical phase, the internal field felt by the muon varies continuously depending on the muon's relative position within the magnetic modulation. This produces a broad, continuous field distribution, which naturally broadens the oscillatory $\mu^+\mathrm{SR}$ signal and leads to increased damping compared with a commensurate AFM state, where muons at the same site in different unit cells experience identical local-field distributions. We therefore treat the reduced fitted asymmetry in the IC phase as a consequence of broad early-time depolarization rather than as evidence for a distinct magnetic volume fraction.

For a purely IC magnetic structure, the spatial modulation of the magnetic moments results in a continuous field distribution, $P(B)$, bound by a maximum field $B_{\text{max}}$. The corresponding muon polarization function is typically described by a zeroth-order Bessel function of the first kind, $J_0(\gamma_\mu B_{\text{max}} t)$ \cite{Yaouanc2011}. A previous high-pressure $\mu^+\mathrm{SR}$ study on doped $\mathrm{Mn}_3\mathrm{Sn}$ fitted the IC helical AFM phase data using a combination of a simple zeroth-order Bessel function and a zeroth-order Bessel function multiplied by a cosine, called a shifted Overhauser distribution \cite{Bhattacharya2024}. We attempted to fit our data using both the pure zeroth-order Bessel function for all muon precession frequencies, as well as the shifted Overhauser distribution model used in the high-pressure study \cite{Bhattacharya2024} and in the related chiral magnet MnGe \cite{Martin2016}. However, this approach yielded very small field distribution widths ($\Delta\nu / \nu_{\text{avg}} < 0.05$). Ultimately, a standard exponentially damped oscillatory function applied to all muon sites with a global phase shift provided the best fit to our experimental data. However, in  absence of microscopic local-field calculations, we treat this phase offset simply as an empirical fitting parameter.

Using the highest-statistics measurement at 5~K, an empirical phase shift of $\phi = 20.3^\circ$ was fitted. The three oscillatory frequencies and their respective asymmetries remain nearly temperature-independent up to 150~K, with the slowest oscillation (AF2) carrying the largest asymmetry. Consequently, the phase shift was kept fixed at $\phi = 20.3^\circ$ for all fits between 5~K and 150~K. 

Above 150~K, the three frequencies undergo a redistribution of spectral weight [Fig.~\ref{fig:ZF_coefs}(a)]. Specifically, $A_{\mathrm{AF2}}$ decreases rapidly from its low-temperature value, while $A_{\mathrm{AF1}}$ and $A_{\mathrm{AF3}}$ increase. At 225~K, the temperature evolution of the precession frequencies, $f_i$, exhibits a kink that deviates from the low-temperature power-law behavior. We attribute this redistribution to the onset of the transition into the commensurate IT-AFM phase. Within this transition regime (175--275~K), $A_{\mathrm{AF2}}$ drops significantly, reaching $\sim 0.007$ at 250~K, whereas $A_{\mathrm{AF1}}$ rises to nearly 0.057. Any change in oscillatory asymmetry expected from impurity phases (e.g., MnO or $\mathrm{Mn}_2\mathrm{Sn}$) becoming paramagnetic would have an asymmetry of approximately 0.003 (see Sec.~\ref{sec:TF}). This is far too small to account for the observed $\sim 0.045$ drop in $A_{\mathrm{AF2}}$. Because the IT-AFM structure is commensurate, its phase is ideally $\phi = 0^\circ$. Because the IT-AFM structure is commensurate, its phase is ideally $\phi = 0^\circ$. To describe the crossover between these coexisting magnetic environments, we empirically scaled the phase down alongside the decreasing AF2 amplitude as $\phi(T) = 20.3^\circ \times A_\mathrm{AF2}(T)/A_\mathrm{AF2}(5~\mathrm{K})$.

\subsubsection{\texorpdfstring{Commensurate IT-AFM phase $\left( T \geq 275~\mathrm{K}\right)$}{Commensurate IT-AFM phase (T >= 275 K)}}

At $T_\mathrm{t} = 275$~K, the AF2 is no longer statistically resolvable (see Appendix~\ref{app:FT}) and the spectra are described using AF1 and AF3.The total fitted amplitude increases across this commensurate region to approximately $A_\mathrm{tot}~\approx~0.20$, consistent with the visually enhanced oscillations in the time spectra [Fig.~\ref{fig:ZF_spectra}]. However, this value remains below the full instrumental asymmetry, indicating that a fraction of the signal is still lost to relaxation faster than what can be resolved in the present measurement. Since this missing fraction persists in the commensurate IT-AFM phase, where the broad field distribution associated with the IC modulation is no longer expected, it points to an additional source of ultra-fast depolarization.

Above 300~K, the fitted amplitudes of the remaining oscillatory components continue to redistribute: $A_{\mathrm{AF1}}$ increases toward 400~K, while $A_{\mathrm{AF3}}$ remains statistically significant at 375~K but is no longer independently resolved at 400~K [Fig.~\ref{fig:ZF_coefs}(a) and Appendix~\ref{app:FT}]. This may arise from thermal expansion of the unit cell changing the electrostatic landscape and making one muon site more favorable. Furthermore, the relaxation rates of the slowly relaxing tail ($\lambda_{\mathrm{tail}}$) and of AF1 increase just below $T_{\text{N}}$, as expected at a magnetic phase transition. Finally, above $T_{\text{N}}$, the non-oscillatory KT component $A_{\mathrm{KT}}$ dominates, as expected for a paramagnetic state where the muon depolarization is dominated by nuclear moments, and all expected instrumental asymmetry is completely recovered.

\section{Discussion}
Before discussing the intrinsic magnetic phases, we first address the possible influence of magnetic impurities. The magnetization measurements reveal a transition and accompanying increase in magnetization at $T_\mathrm{im} = 200$~K, which has previously been attributed to a $\mathrm{Mn}_{2-x}\mathrm{Sn}$ impurity phase~\cite{Yano2024}. While bulk $\mathrm{Mn}_2\mathrm{Sn}$ is ferrimagnetic with a Curie temperature $T_\text{C} \approx 250$~K~\cite{Fuglsby2015}, this transition is highly sensitive to Mn deficiency and can shift to lower temperatures~\cite{Xu2013}. By comparing the magnitude of the magnetization step ($\Delta M$) across $T_\mathrm{im}$ with the reference magnetization of $\mathrm{Mn}_2\mathrm{Sn}$ ($\approx 25$~emu/g), we estimate the impurity fraction to be only $\sim 1\%$.

The small TF-$\mu^+\mathrm{SR}$ recovery between 100~K and 200~K, corresponding to approximately 1--2\% of the total asymmetry, is compatible with the impurity estimate and could contain contributions from $\mathrm{Mn}_{2-x}\mathrm{Sn}$ and trace MnO from surface oxidation of the powder (AFM with $T_\mathrm{N} = 120$~K~\cite{Shull1951}). A second TF recovery of comparable size is observed above $\sim 250$~K. Since the bulk $M(T)$ data do not show a corresponding sharp ferromagnetic or ferrimagnetic anomaly in this temperature range, we do not attribute this second feature to a FM/ferrimagnetic impurity. Instead, we regard it as most likely connected to the intrinsic IC-helical to IT-AFM reconfiguration of $\mathrm{Mn}_3\mathrm{Sn}$, although a very small AFM minority contribution cannot be excluded. The magnetic reconfiguration into a more symmetric structure may modify the internal-field distribution and cause partial field cancellation at a minority stopping site, analogous to behavior observed in the hexagonal AFM FeSn~\cite{Hartmann1987}. Alternatively, given its small magnitude, the shift could originate from minor, temperature-dependent variations in $\alpha$ due to measurements over a large temperature range.

Regardless of its exact origin, treating this possible impurity-related asymmetry as a conservative upper bound shows it is far too small to account for the main ZF-$\mu^+\mathrm{SR}$ observations. In the IC-helical/IT-AFM crossover regime, the redistribution of oscillatory spectral weight is of order $\Delta A \sim 0.045$, whereas the impurity contribution inferred from TF-$\mu^+\mathrm{SR}$ would correspond to only $\sim 0.003$ in the ZF oscillatory asymmetry. We therefore conclude that the ZF-$\mu^+\mathrm{SR}$ spectra predominantly probe the intrinsic magnetic evolution of the main $\mathrm{Mn}_3\mathrm{Sn}$ phase.

From our magnetization measurements, we identify a low-temperature transition at $T_\mathrm{f} = 21$~K. The strong bifurcation between ZFC and FC curves below $T_\mathrm{f}$ is widely reported and often attributed to a partial FM spin-glass phase \cite{Feng2006, Song2020_ComplicatedMagStruct}. Indeed, our magnetization data exhibit a clear FM signature, with a continuous increase in both the total magnetization and the hysteresis as the temperature is lowered. However, although our AC magnetic susceptibility (ACMS) data reveal a distinct peak at $T_\mathrm{f} = 21$~K, we do not observe the characteristic frequency-dependent shift in the peak position of $\chi'$ in ACMS, which is expected for a spin-glass transition \cite{Tholence1980}.

Furthermore, our ZF $\mu^+\mathrm{SR}$ spectra lack the signature of a statically frozen spin glass characterized by a static KT polarization function \cite{Hayano1979, Uemura1985}. Although it has been speculated that the glassy phase in $\text{Mn}_3\text{Sn}$ is driven by localized magnetic clusters of excess Mn occupying Sn sites \cite{Feng2006, Ikhlas2020}, such a phase, even as a minority volume fraction, should result in a resolvable static component in our $\mu^+\mathrm{SR}$ spectra.

Instead, upon cooling through $T_\mathrm{f}$, we observe a distinct peak in the relaxation rate $\lambda_{\mathrm{AF1}}$. This peak in relaxation indicates a phase transition consistent with the features observed in our magnetization measurements. However, rather than static spin-glass freezing, we observe a distinct slowing of spin fluctuations. This suggests that for our $\mathrm{Mn}_{2.99}\mathrm{Sn}$ sample, the intrinsic transition is distinctly separate from the static glassy freezing associated with Mn impurities in some other $\mathrm{Mn}_3\mathrm{Sn}$ samples \cite{Ikhlas2020, Feng2006}. Consequently, the FM phase appears to be decoupled from the glassy behavior, though the nature of the evolving spin dynamics requires further investigation.

In the IC helical phase, we found that a regular damped cosine with an empirical phase shift of $\phi = 20.3^\circ$ provides a significantly better fit than the simple Overhauser or zeroth-order Bessel functions used in previous studies~\cite{Bhattacharya2024}. Because the symmetric Bessel functions rely on ideal muon sampling of the field distribution, their failure does not independently prove an asymmetric internal-field distribution. However, this asymmetric fit is highly consistent with the complex, amplitude-modulated order established by recent neutron and X-ray scattering studies on $\mathrm{Mn}_3\mathrm{Sn}$~\cite{Cable1993,Park2018,Wang2023multiK,Chen2024_IntertwinedCDWSDW}.

In the helical phase, we observe three oscillation frequencies, in contrast to the two reported in the high-pressure experiment by Bhattacharya \textit{et al.}~\cite{Bhattacharya2024}. The additional weak component may be visible here because of the longer counting time and the absence of pressure-cell background. The presence of three frequencies aligns with the three stopping sites predicted by DFT calculations by Bhattacharya \textit{et al.}~\cite{Bhattacharya2024}. As the system is heated beyond 150~K, we observe a continuous redistribution of spectral weight between these oscillations. Approaching the IT-AFM phase, the initially faint frequency, AF3, grows in amplitude, while the primary low-temperature frequency, AF2, diminishes in spectral weight until it is no longer resolvable at the transition temperature, $T_{\mathrm{t}} \approx 275$~K [Fig.~\ref{fig:ZF_coefs}(a)]. 

 We interpret this spectral redistribution as evidence for coexisting local magnetic environments associated with the IC helical and IT-AFM states. Specifically, in the context of an amplitude-varying spin-density wave (SDW), the diminishing spectral weight of the AF2 component is consistent with the gradual suppression of the IC helical phase in favor of the commensurate IT-AFM phase. Previous studies have suggested a rotation of the easy axis from the helical $[11\bar{2}0]$ direction to the IT-AFM $[0001]$ direction prior to the bulk transition at $T_{\mathrm{t}}$~\cite{Duan2015, Sung2018}. While these reports attribute this shift to an internal transition within the IC helical phase, our results instead suggest that this easy-axis shift tracks the continuous magnetic crossover into the IT-AFM state. This is supported by detailed neutron powder diffraction, which establishes a broad temperature window (spanning roughly 60 K) where the commensurate and incommensurate magnetic reflections gradually transition into one another~\cite{Song2020_ComplicatedMagStruct}. Thus, while we do not completely rule out a macroscopic phase separation, their observed behaviour closely parallels the continuous crossover observed in our measurements.

At and above 275~K, the transition into the IT-AFM phase is completed, marked by the no longer resolvable AF2 component. This transition is further visible in our magnetization data through an increase in magnetization above $T_\mathrm{t} = 275$~K, which is characteristic of the weak FM moment of the IT-AFM structure~\cite{Mn3Sn_struc1, Mn3Sn_struc2}. It is useful to distinguish the reduced fitted asymmetry in the IC-helical phase from the persistent missing fraction observed in the commensurate IT-AFM phase. In the IC-helical phase, the spatial modulation of the ordered moments naturally produces a broad distribution of internal fields, and part of the early-time depolarization may therefore become difficult to resolve within the instrumental time window. Thus, this reduction in observable asymmetry should not be misinterpreted as phase separation or a distinct magnetic volume fraction. By contrast, above $T_\mathrm{t}$ the IC modulation is absent and the magnetic structure is commensurate; the persistence of a reduced resolvable initial asymmetry therefore suggests that a subset of muons still experiences very broad or very large local fields that depolarize the signal within the first few nanoseconds.

One possible microscopic origin of this unresolved depolarization is a minority population of muons stopping in regions where the magnetic texture varies strongly on the nanoscale. In the IT-AFM phase, the degeneracy of the spin configurations allows $\mathrm{Mn}_3\mathrm{Sn}$ to form multiple magnetic domains~\cite{Jacobsen_2026,Reichlova2019,Li2019,Tsukamoto_domainwalls}, and the associated domain walls have been discussed as topological or chiral spin textures with a continuous rotation of the local magnetic order~\cite{Li2019,Tsukamoto_domainwalls}. Muons stopping in or near such domain-wall regions could therefore experience a broad distribution of local fields, producing relaxation that is too fast to resolve in the present time window.

However, this interpretation should be regarded as tentative. Other sources of ultrafast depolarization, such as grain-boundary strain in the powder, unresolved minority muon stopping sites with very large local fields, or local disorder near magnetic/domain boundaries, could produce a similar missing fraction. Thus, we do not directly equate the missing asymmetry with a domain-wall volume fraction. Instead, we interpret it more generally as evidence for a minority subset of muons experiencing an unresolved, highly inhomogeneous local magnetic environment. Longitudinal-field $\mu^+\mathrm{SR}$, ideally on single crystals, together with complementary spatial probes such as small-angle neutron scattering (SANS) or magnetic imaging, would be needed to determine whether domain walls are the dominant source of this missing fraction~\cite{Kreyssig2009}.

Lastly, a second redistribution occurs above 325~K with a decrease in $A_{\mathrm{AF3}}$, which we tentatively attribute to thermal expansion of the unit cell making the AF1 muon site more favorable. Further DFT calculations of the muon sites, utilizing precise temperature-resolved crystal structures as inputs, are needed to confirm this.

\section{Conclusion}
Our combined $\mu^+\mathrm{SR}$ and bulk magnetization measurements on $\mathrm{Mn}_{2.99}\mathrm{Sn}$ reveal a dynamic magnetic landscape across the three primary bulk transitions: $T_\mathrm{f} = 21$~K, $T_\mathrm{t} \approx 275$~K, and the N\'eel temperature $T_\mathrm{N} = 418$~K. 

Below $T_\mathrm{f}$, we do not observe any signature of a frozen spin glass state in this sample. However, we still observe an increasing FM component accompanied by a distinct slowing down of spin fluctuations. This indicates that the low-temperature FM signatures might be intrinsic to the main crystalline phase and decoupled from the static glassy freezing observed in other $\mathrm{Mn}_3\mathrm{Sn}$ samples. By demonstrating that the FM transition can appear independently of static glassiness, these results shed new light on a regime previously interpreted as a coupled, partial FM spin-glass state.

Within the IC helical AFM regime, damped oscillations with an empirical common phase offset provide the best fit to our measured ZF spectra, which iscompatible with the anharmonic magnetic order established by recent scattering studies~\cite{Wang2023multiK,Chen2024_IntertwinedCDWSDW}.Upon heating, we identify a broad magnetic crossover between the IC helical and IT-AFM phases. The continuous redistribution of spectral weight between the oscillations tracks this crossover, suggesting that previously reported easy-axis rotations are a direct consequence of this gradual magnetic transition.

In the IT-AFM state above $T_\mathrm{t}$, a persistent missing fraction in the initial asymmetry indicates that a subset of implanted muons, corresponding to roughly 20\% of the sample-related asymmetry, undergoes unresolved ultrafast depolarization. The microscopic origin of this missing fraction remains open. Possible contributions include domain-wall regions, grain-boundary or disorder effects in the powder, and unresolved minority muon stopping environments. Longitudinal-field $\mu^+\mathrm{SR}$ on single crystals, ideally combined with complementary spatial probes, will be required to distinguish between these possibilities. Additionally, we observe a further redistribution of muon sites above 325~K, which we tentatively attribute to thermal expansion favoring a specific muon site.

Ultimately, our results reveal a highly dynamic magnetic landscape in $\mathrm{Mn}_3\mathrm{Sn}$. By utilizing the local probe of $\mu^+\mathrm{SR}$, we show how distinct local magnetic environments coexist over a large temperature range and evolve continuously with temperature. To fully understand this redistribution and the local coexistence of the helical and IT-AFM environments, further DFT calculations utilizing precise temperature-resolved crystal structures, alongside detailed neutron diffraction studies, are needed.

\bigskip
\noindent{\bf \large{Acknowledgments}}
\vspace{4pt}

\noindent
We gratefully acknowledge S.~Jana for the sample synthesis.

\bigskip
\noindent{\bf \large{Funding Information}}
\vspace{4pt}

\noindent
U.M. acknowledges funding from the KTH-SCI doctoral excellence program. O.K.F. is supported by the Swedish Research Council (VR) through Grant 2022-06217, ÅForsk repatriation grant (ref.nr 25-1) and the Ruth and Nils-Erik Stenbäck Foundation.

\bigskip

\bigskip
\noindent{\bf \large{Data Availability and Analysis}}
\vspace{4pt}
\noindent
Data and their analysis are available from the corresponding authors upon reasonable request.

\bigskip
\hrule     
\bigskip

\setcounter{table}{0}
\setcounter{figure}{0}
\setcounter{equation}{0}
\renewcommand{\thetable}{S\arabic{table}}
\renewcommand{\thefigure}{S\arabic{figure}}
\renewcommand{\theequation}{S\arabic{equation}}

\appendix

\section{Number of frequencies in zero-field}\label{app:FT}
\begin{figure}[htpb]
    \centering
    \includegraphics[width=1\linewidth]{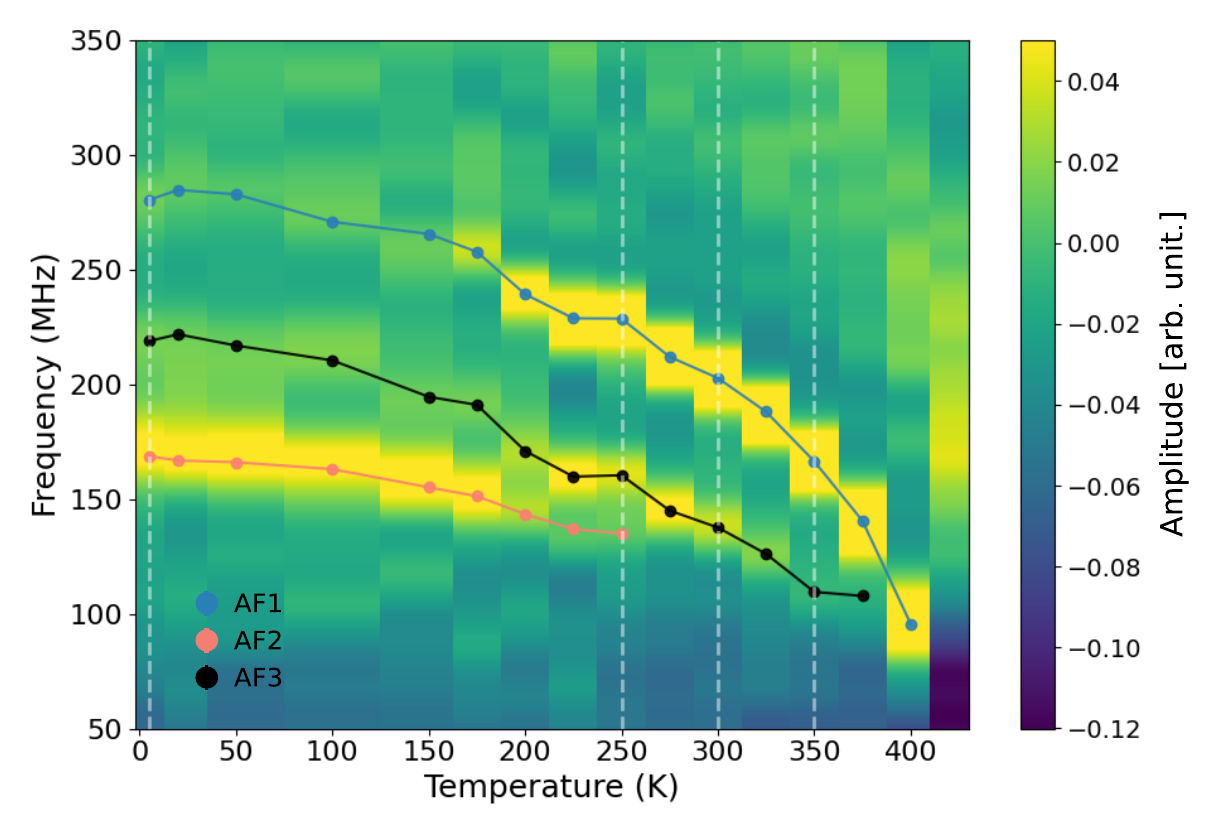}
    \caption{\textbf{Fourier transform heat map of the ZF-$\boldsymbol{\mu}^+$\textbf{SR} signal.} 2D heat map showing the real part of the Fourier transform (FT) of the asymmetry signal over the first $0.05~\mu\mathrm{s}$ as a function of temperature. Vertical dashed lines indicate temperatures where high-statistics measurements were performed. Overlaid points denote the fitted zero-field (ZF) precession frequencies (AF1, AF2, and AF3) extracted using Eq.~\eqref{eq:ZF}, matching the parameters presented in Fig.~\ref{fig:ZF_coefs}(a).}
    \label{fig:heat_map}
\end{figure}

Fig.~\ref{fig:heat_map} shows the real part of the Fourier transform of the $\mu^+\mathrm{SR}$ ZF spectra, overlaid with the fitted ZF frequencies. The dashed lines denote high-statistics (HS) measurements (taken at 5~K, 250~K, 300~K, and 350~K), which have approximately double the standard statistics. Consequently, the noise floor outside of the three fitted frequency bands is visibly lower in these HS datasets. 

Because some frequency components become very weak with temperature, it is important to justify statistically the number of independently resolvable components used in the fits. As summarized in Table~\ref{tab:f_test_results}, we first confirm that adding AF3 to a two-frequency baseline at 5~K produces a statistically significant improvement. We then apply the same analysis where AF2 becomes weak near $T_\mathrm{t}$ (250~K and 275~K) and where AF3 becomes weak on approaching 400~K (350~K, 375~K, and 400~K). The added-component amplitude is left unconstrained in these tests. At 275~K, the $F$-test gives $p = 0.092$, so AF2 is no longer statistically significant ($p > 0.05$) and is excluded from the final model at 275~K and above. This establishes loss of independent resolvability, not a frequency that falls to zero. At 375~K, AF3 remains statistically significant, whereas at 400~K no stable positive AF3 amplitude can be obtained; the single-frequency model is therefore used at 400~K, and that dataset is not included in Table~\ref{tab:f_test_results}.

Because $\mu^+\mathrm{SR}$ time spectra consist of many bins, our fits have a very large number of degrees of freedom ($\text{ndf} \approx 40,000$). Here, simply observing a drop in the absolute $\chi^2$ value is insufficient to justify adding more parameters, as any additional parameter will mathematically reduce the $\chi^2$. To test whether the improvement represents a true physical signal rather than just fitting statistical noise, we utilized an $F$-test for nested models. The $F$-statistic is calculated as follows:
\begin{equation}
    F = \frac{(\chi^2_1 - \chi^2_2) / (\Delta\text{ndf})}{\chi^2_2 / \text{ndf}_2}
    \label{eq:f_test}
\end{equation}
where $\chi^2_1$ and $\text{ndf}_1$ correspond to the simpler model (fewer frequencies), and $\chi^2_2$ and $\text{ndf}_2$ correspond to the more complex model. The $F$-test inherently penalizes overfitting by requiring the normalized drop in $\chi^2$ (numerator) to significantly exceed the baseline statistical noise of the fit (denominator). The $p$-value is then calculated by evaluating the survival function (the right-tail probability) of the $F$-distribution at the observed $F$-statistic, using $\Delta\text{ndf}$ and $\text{ndf}_2$ as the degrees of freedom. A resulting $p$-value $< 0.05$ indicates, within this nested-model comparison, that including the additional oscillatory component produces a statistically significant improvement in the fit.

\begin{table*}[t]
    \centering
    \caption{Summary of the $F$-test for nested models used to justify the inclusion of additional frequency components at various temperatures. $\chi^2_1$ and $\chi^2_2$ represent the absolute chi-squared values for the models with fewer and more frequencies, respectively, alongside their degrees of freedom (ndf). A $p$-value $< 0.05$ indicates that the inclusion of the additional frequencies is statistically significant. The unconstrained fitted asymmetry of the extra frequency is also provided.}
    \label{tab:f_test_results}
    \begin{tabular}{llccccc}
        \hline\hline
        Temp & Comparison & $\chi^2_1$ (ndf$_1$) & $\chi^2_2$ (ndf$_2$) & $F$-statistic & $p$-value & Fitted Asym. \\
        \hline
        5 K  & 2 vs.\ 3 Freq & 40816.4 (40930) & 40772.8 (40927) & 14.59 & $1.71\cdot10^{-9}$ & $A_\mathrm{AF3} = 0.0070(15)$ \\
        250 K & 2 vs.\ 3 Freq & 41055.7 (40931) & 41033.2 (40928) & 7.48 & $5.29\cdot10^{-5}$ & $A_\mathrm{AF2} = 0.0070(26)$ \\
        275 K & 2 vs.\ 3 Freq & 41350.9 (40931) & 41344.4 (40928) & 2.14 & $0.092$ & $A_\mathrm{AF2} = 0.00124(94)$ \\
        350 K & 1 vs.\ 2 Freq & 41197.8 (40934) & 41182.1 (40931) & 5.20 & $0.00136$ & $A_\mathrm{AF3} = 0.039(12)$ \\
        375 K & 1 vs.\ 2 Freq & 41239.2 (40934) & 41228.0 (40931) &  3.71 & $0.0111$ & $A_\mathrm{AF3} = 0.034(22)$ \\
        \hline\hline
    \end{tabular}
\end{table*}

\section{Magnetization}\label{SI:mag}
\begin{figure*}
    \centering
    \includegraphics[width=1\linewidth]{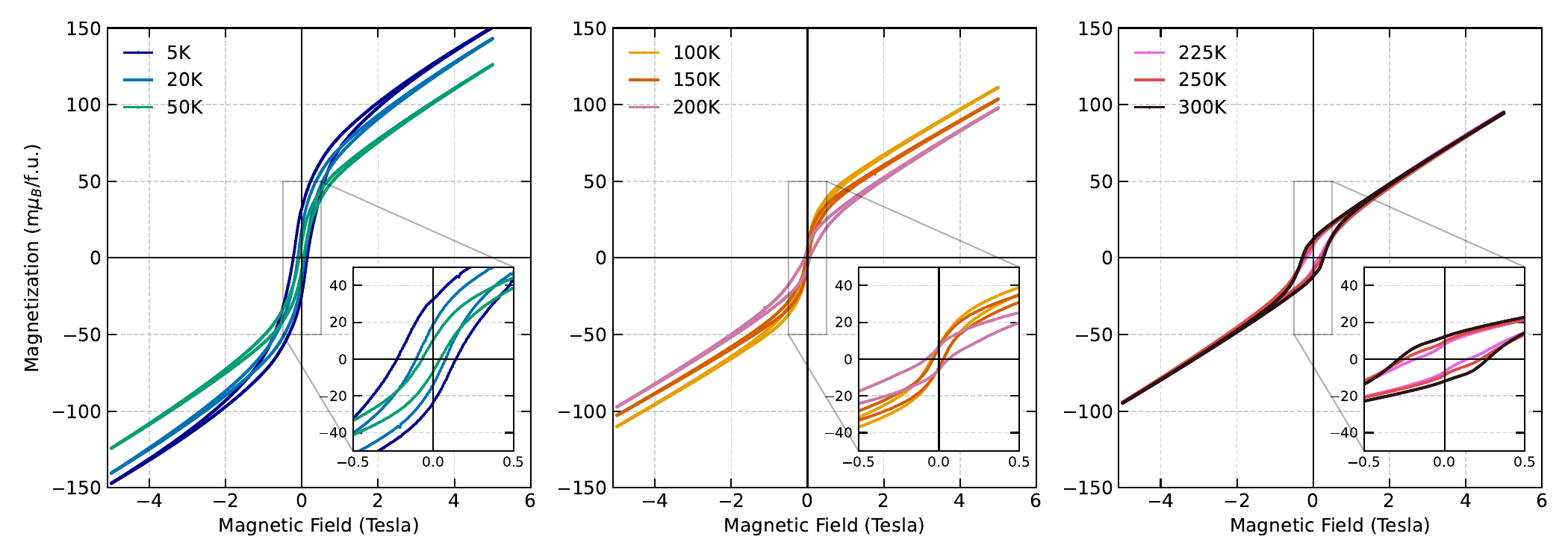}
    \caption{\textbf{Extended isothermal bulk magnetization.} Magnetization of $\mathrm{Mn}_3\mathrm{Sn}$ as a function of applied external magnetic field, measured across a wide range of temperatures. The data are grouped into three panels by temperature range for clarity. The insets provide a magnified view of the low-field region to more clearly illustrate the temperature evolution of the hysteresis loops.}
    \label{fig:MH_all_temps}
\end{figure*}

Fig.~\ref{fig:MH_all_temps} shows the magnetization curves for temperatures spanning from 5~K to 300~K. At 300~K, within the inverse triangular (IT) AFM phase, the material exhibits a relatively large hysteresis alongside a damped total magnetic response. This behavior is consistent with the canted IT-AFM structure present in this regime~\cite{Mn3Sn_struc1,Sandratskii1996_WeakFM,Kren1975_Mn3SnPhaseTrans,Mn3Sn_struc2,Tomiyoshi1982_MagStructWeakFM, Nagamiya1982}. Cooling through the bulk transition temperature $T_\mathrm{t} = 275$~K and into the helical phase, the hysteresis narrows visibly at 225~K, although the total magnetization remains comparable. Below 200~K, it is also important to consider the potential background contribution of a ferrimagnetic Mn$_{2-x}$Sn impurity phase. As the temperature is lowered past 200~K, the $M(H)$ curves lose their sharp features and adopt a smoother, S-shaped profile with minimal hysteresis. This smooth S-shape persists down to 50~K. However, as the temperature drops below 50~K, the hysteresis loop begins to broaden significantly once again. This distinct re-emergence of the coercive field at low temperatures indicates a growing ferromagnetic component in the main $\mathrm{Mn}_3\mathrm{Sn}$ phase.

\clearpage 

\bibliography{Refs_2} 
\end{document}